\documentclass[
reprint,
prd,
amsmath,amssymb,aps,
nofootinbib
]{revtex4-2}
\usepackage{graphicx}
\usepackage{dcolumn}
\usepackage{bm}
\usepackage{amsmath}
\usepackage[dvipsnames]{xcolor}
\usepackage{hyperref}
\usepackage{booktabs}
\usepackage{comment}

\makeatletter

\renewcommand\onecolumngrid{%
  \do@columngrid{one}{\@ne}%
  \def\set@footnotewidth{\onecolumngrid}%
  \def\footnoterule{\kern-6pt\hrule width 1.5in\kern6pt}%
}

\renewcommand\twocolumngrid{%
  \def\footnoterule{%
    \dimen@\skip\footins\divide\dimen@\thr@@
    \kern-\dimen@\hrule width.5in\kern\dimen@}%
  \do@columngrid{mlt}{\tw@}%
}

\makeatother

\def \HH {\mathcal{H}}

\newcommand{\dm}{{\rm dm}}
\newcommand{\m}{{\rm m}}
\newcommand{\g}{{\rm g}}
\newcommand{\ba}{{\rm b}}
\newcommand{\de}{{\rm de}}
\newcommand{\ep}{\mathcal{E}}
\newcommand{\Z}{\mathcal{Z}}
\newcommand{\teststress}{\mathcal{N}}
\newcommand{\testeuler}{\mathcal{N}^{\rm Euler}}
\newcommand{\testzi}{\mathcal{N}^{0i}}
\newcommand{\testij}{\mathcal{N}^{\rm slip}}

\providecommand\mnras{Mon.\ Not.\ Roy.\ Astron.\ Soc. }
\providecommand\aap{Astron.\ Astrophys. }
\providecommand\prd{Phys.\ Rev.\ D} 
\providecommand\prl{Phys.\ Rev.\ Lett. }

\providecommand\jcap{JCAP }

\providecommand\apj{Astrophys.\ J. }

\begin{document}

\title{Using the 0i Einstein equation to break dark-sector degeneracies}

\author{Camille Bonvin}
\email{camille.bonvin@unige.ch}
\affiliation{%
D\'epartement de Physique Th\'eorique and Center for Astroparticle Physics, Universit\'e de Gen\`eve, Quai E. Ansermet 24, CH-1211 Geneva 4, Switzerland}%

\begin{abstract}
Over the past decade, cosmological observations have started to challenge the $\Lambda$CDM model. If these tensions persist, they immediately raise one fundamental question: what new physics could be responsible for them? More specifically, do these deviations arise from a modification of gravity, from an evolving or clustering dark energy component, or from dark matter properties beyond the cold dark matter paradigm?
In this paper we show that, with current methods, any observed deviation can always be attributed to more than one sector, preventing us from determining if the tension is due to gravity, dark energy or dark matter. The main reason is that current tests rely on unobservable quantities. These quantities are then either related to observables by assuming the validity of specific equations, or compared with theoretical predictions computed within a given class of theories. Both approaches inevitably break the model independence of the tests, leading to degeneracies between gravity, dark energy, and dark matter. To overcome this limitation, we construct a set of three tests that rely only on observable quantities and that, when combined, can distinguish between the three sectors. We show that the $0i$ Einstein equation -- which is never tested in standard analyses -- adds crucial information to that provided by the gravitational slip equation and the Euler equation.

\end{abstract}

\maketitle

\section{Introduction}\label{sec:intro}

One of the main goal of current and upcoming cosmological surveys is to test the validity of the $\Lambda$CDM model. In particular, measurements of the expansion rate of the Universe and the growth rate of structure, are used to probe the mechanism responsible for cosmic acceleration, i.e.\ to understand if the acceleration is due to a cosmological constant, an evolving dark energy, or a modification of gravity. In parallel, cosmological observations allow us to test the cold dark matter (CDM) paradigm by searching for additional interactions in the dark sector and deviations from perfect-fluid behavior. In the past decades, a plethora of analyses have been performed, placing constraints on all kind of extensions of the $\Lambda$CDM model, see e.g.,~\cite{Planck:2018vyg,eBOSS:2020yzd,DES:2022ccp,DESI:2025zgx}. These analyses of alternative models are particularly relevant in view of the tensions between data sets -- Cosmic Microwave Background, Baryon Acoustic Oscillations + supernovae distances, gravitational lensing and galaxy clustering -- that have emerged in the past decade and that tend to favour some of the beyond-$\Lambda$CDM models~\cite{Tutusaus:2023aux,DESI:2025zgx,Ye:2024ywg,Chudaykin:2024gol,Akarsu:2024eoo,Pan:2025psn,Wolf:2024stt,Wolf:2025acj,Wolf:2025jed,vanderWesthuizen:2025rip,Ye:2026yqk,Pompeu:2026zad}.    

In this paper, we show that although these analyses can robustly constrain individual models beyond $\Lambda$CDM, they are unable to uniquely connect an observed feature to a specific class of theories. At the background level this has long been known: the expansion history alone cannot distinguish between the three classes of theories beyond $\Lambda$CDM, i.e.\ between modified theories of gravity, dark energy and extended dark matter models. There is indeed sufficient freedom in each of these classes to reproduce essentially any desired expansion history. A striking example of this degeneracy is the phantom crossing observed by DESI~\cite{DESI:2025zgx}, which can be equally well reproduced by clustering dark energy~\cite{Creminelli:2008wc}, modifications of gravity~\cite{Ye:2024ywg,Chudaykin:2024gol,Pan:2025psn,Wolf:2025jed,Wolf:2024stt} or interacting dark matter models~\cite{Pompeu:2026zad,Clini:2026kdj}. For this reason, the community has increasingly advocated using the growth of perturbations to break degeneracies between modified gravity, dark energy and extended dark matter models, see e.g.~\cite{Wang:2007ht,Zhao:2009fn,Zhao:2010dz,Simpson:2010yt,EuclidTheoryWorkingGroup:2012gxx,Huterer:2013xky}. 

In this work, we show, however, that the perturbed quantities that we can currently measure do not contain enough information to discriminate between the three classes. Current surveys can indeed measure only two perturbed quantities: the galaxy velocities $V_\g$ measured from redshift-space distortions, and the Weyl potential $\Psi_W=(\Phi+\Psi)/2$ measured through gravitational lensing and the Integrated Sachs-Wolfe effect (here $\Psi$ and $\Phi$ denote the temporal and spatial distortions of the metric, respectively). The matter density fluctuation, $\delta$, cannot be measured from galaxy clustering observations since it is fully degenerate with the unknown galaxy bias. The key limitation, then, is that $V_\g$ and $\Psi_W$ are affected in a similar way by modified gravity, dark energy and extended dark matter, making it impossible to distinguish between the three classes of theories using these measurements alone. Improving the precision of measurements of $V_\g$ and $\Psi_W$ will not resolve this degeneracy, since one can always construct models within each of the three classes that reproduce the observations equally well. Moreover, testing the relation between $V_\g$ and $\Psi_W$, as done for example with the $E_G$ statistic~\cite{Zhang:2007nk,Alam:2016qcl,Amon2018MNRAS.479.3422A, Reyes:2010tr, Blake:2016fcm, Pullen2016, MoradinezhadDizgah:2016pqy, Ghosh:2018ijm, Wenzl:2024sug, Abidi_2023_10419,Grimm:2024fui}, is also unable to break the degeneracy, since all three classes of theories can modify this relation in indistinguishable ways. What makes these degeneracies particularly insidious is that they affect not only modified gravity and dark energy, the two classes of theories developed to explain the accelerated expansion of the Universe, but also extended dark matter models, which are motivated by entirely different physical considerations, yet turn out to leave the same signatures on the evolution of $\Psi_W$ and $V_\g$.

A game-changer, first identified in~\cite{Castello:2022uuu,Bonvin:2022tii},  is the gravitational redshift effect. This effect is sensitive to the time distortion $\Psi$ alone, and will be measurable by the newest generation of surveys~\cite{Breton:2018wzk,Beutler:2020evf,Saga:2021jrh,Sobral-Blanco:2022oel, Bonvin:2023jjq}. It will allow us to test the validity of the Euler equation and the gravitational slip equation (the $i\neq j$ Einstein equation), thereby providing a way to distinguish modified gravity from dark matter models with additional interactions~\cite{Castello:2024jmq,Castello:2024lhl}. In this work we build on this argument and show that distinguishing among all three classes of theories -- modified gravity, dark energy and extended dark matter -- requires testing an additional equation: the $0i$ Einstein equation.

This equation is systematically omitted in standard tests of gravity. Indeed, owing to Bianchi identities~\cite{Wald:1984rg}, all information is contained in the Poisson equation, the gravitational slip equation, the Euler equation and the continuity equation. The $0i$ Einstein equation is therefore redundant and provides no additional information. We will show, however, that when the goal is not to test a specific class of theories but to distinguish among different classes, the $0i$ Einstein equation plays a central role and should replace the Poisson equation. The reason is that the Poisson equation depends on the density fluctuation $\delta$, which is not directly observable. It therefore cannot be tested in a model-independent way and, as a result, cannot distinguish among the three classes.

In this paper, we construct a set of three tests based on the relevant equations: $\testij$, which probes the gravitational slip equation; $\testzi$, which probes the $0i$ Einstein equation; and $\testeuler$, which probes the Euler equation. These tests depend only on measurable quantities, and we will show that, when combined, they can unambiguously distinguish among modified gravity, dark energy and extended dark matter.

The rest of the paper is organized as follows. In Section~\ref{sec:level0}, we explicitly demonstrate the degeneracy among the three classes of theories. We derive the evolution equations at linear order in perturbation theory for five specific models drawn from the modified gravity, dark energy, and dark matter classes, and show that current methods fail to distinguish among them. In Section~\ref{sec:level1}, we demonstrate that testing the gravitational slip equation using $V_\g$ and $\Psi_W$, as commonly done in cosmological analyses, also fails to break these degeneracies. In Section~\ref{sec:level2}, we present the three tests and show that, when combined, they can distinguish among the three classes. Because the Poisson equation is not tested within our framework, we show in Section~\ref{sec:Poisson} that our approach does not discard any information. Finally, we conclude in Section~\ref{sec:conclusion}.

\section{Level zero: Testing departures from $\Lambda$CDM with the Weyl potential and galaxy velocities}
\label{sec:level0}

We start by showing why current methods cannot distinguish between the three classes. The argument is split into two parts: \textit{Level zero} in this section, where we concentrate on standard analyses that compare the measurements of $V_\g$ and $\Psi_W$ with theoretical predictions in a given class of theories; and \textit{Level one} in Section~\ref{sec:level1}, where we look at analyses that combine $V_\g$ and $\Psi_W$ to test the gravitational slip equation.

To illustrate the degeneracies, we explore five representative models that belong to the three classes: modified gravity, dark matter and dark energy. As an example of the class of modified gravity we consider Generalized Brans-Dicke (GBD) theories. In the dark matter class we explore two types of models beyond CDM: coupled quintessence (CQ), where dark matter interacts with dark energy; and viscous dark matter, where dark matter does not behave as a perfect fluid. Finally, in the dark energy class we also explore two models: an evolving dark energy model, where a smooth quintessence field leads to an equation of state $w_\de\neq -1$, and a clustering dark energy model where the quintessence field acquires non-negligible perturbations. These five models have been chosen to capture the fundamental features of the three classes, and we will see that the results that we obtain are much more general and apply to the vast majority of models in the three classes. We discuss this in more detail in Section~\ref{sec:level2}.
We start by presenting the five models and then show their degeneracies.

\subsection{The five examples}

In all examples, we work with a flat\footnote{Similar tests could be built in a curved spacetime, constructed from the same combination of observables. However, given the tight constraints on the curvature from current observations~\cite{Planck:2018vyg}, the deviations from flatness in the tests are expected to be marginal.} perturbed Friedmann-Lema\^itre-Robertson-Walker (FLRW) metric
\begin{align}
\mathrm ds^2=a^2(\tau)\left[-(1+2\Psi(\mathbf{x},\tau))\mathrm d\tau^2+ (1-2\Phi(\mathbf{x},\tau))\mathrm d\mathbf{x}^2\right]\,,\nonumber
\end{align}
where $a$ is the scale factor which depends on conformal time $\tau$, and $\Psi$ and $\Phi$ are the two gravitational potentials. 

\subsubsection{Modified gravity: Generalized Brans-Dicke theories (GBD)}

GBD theories contain a scalar field $\varphi$ non-minimally coupled to matter, whose Lagrangian takes the form
\begin{align}
S^{\rm GBD}=\int d^4 x &\sqrt{-g}\Bigg[\frac{A^{-2}(\varphi)R}{16\pi G}-\frac{1}{2}\nabla_\mu\varphi\nabla^\mu\varphi \label{eq:SGBD}\\
&-U(\varphi)+\mathcal{L}_{\rm m}(\psi_{\rm sm},\psi_{\rm dm},g_{\mu\nu})\Bigg] \, .  \nonumber
\end{align}
Here $G$ is the Newton's constant, $R$ is the Ricci scalar, $g$ is the determinant of the metric $g_{\mu\nu}$, $A(\varphi)$ is a generic function of $\varphi$ encoding the coupling to matter and $U(\varphi)$ is the scalar field potential. $\mathcal{L}_{\rm}(\psi_{\rm sm},\psi_{\rm dm},g_{\mu\nu})$ is the Lagrangian density of the matter which contains the Standard Model (sm) particle fields, collectively denoted as $\psi_{\rm sm}$, and the dark matter (dm) particles, denoted as $\psi_{\rm dm}$. Both Standard Model and dark matter particles follow the geodesics of the metric $g_{\mu\nu}$. 

The linear perturbation equations that relate the metric perturbations to the matter density fluctuations and velocities are derived in Appendix~\ref{app:GBD}. We see that in GBD theories the Poisson, gravitational slip and $0i$ Einstein equations are all modified with respect to $\Lambda$CDM, while the Euler and continuity equations keep the standard form. The strength of the coupling to the scalar field is encoded in the parameter $
\beta=A_{,\varphi}/A$, where $A_{,\varphi}$ denotes the derivative of $A$ with respect to $\varphi$.

\subsubsection{Dark matter with a fifth force: Coupled Quintessence (CQ)}

In CQ models, dark matter is coupled to a scalar field $\varphi$, which plays the role of dark energy. The action in this case takes the form 
\begin{align}
S^{\rm CQ}=\int d^4 x &\sqrt{-g}\Bigg[\frac{R}{16\pi G}-\frac{1}{2}\nabla_\mu\varphi\nabla^\mu\varphi-U(\varphi)\label{eq:SCQ}\\
&+\mathcal{L}_{\rm sm}(\psi_{\rm sm}, g_{\mu\nu})+\mathcal{L}_{\rm dm}(\psi_{\rm dm}, A^2(\varphi)g_{\mu\nu})\Bigg] \, . \nonumber  
\end{align}
Here $g_{\mu\nu}$ denotes the metric of the ``baryon frame'', i.e.\ the metric whose geodesics are followed by the Standard Model particles. Dark matter particles on the other hand follow geodesics of the metric $A^2(\varphi)g_{\mu\nu}$.

The linear perturbation equations are written in Appendix~\ref{app:CQ}. We see that in CQ models the Poisson and gravitational slip equations are not modified. The $0i$ Einstein equation is affected by the flow of the scalar field. The continuity equations for baryons and dark matter remain standard, since there is no energy exchange between dark matter and dark energy, but the Euler equation for dark matter is modified by the coupling. As in GBD theories, the strength of the coupling is encoded in $\beta=A_{,\varphi}/A$, however, in CQ models, $\beta$ only affects dark matter particles and does not impact Standard Model particles.

\subsubsection{Viscous dark matter}

In viscous dark matter models, contrary to CQ models, there are no new degrees of freedom coupled to dark matter. These models instead differ from CDM in the form of their energy–momentum tensor: rather than behaving as a perfect fluid, the dark matter exhibits viscous stresses that may arise from a variety of microscopic processes. Following Eckart's theory of relativistic fluids~\cite{Eckart:1940te}, the energy-momentum tensor of dark matter can be written as
\begin{align}
T^{\mu}_{\dm\,\nu}&=\rho_\dm u^\mu_\dm u_{\dm\,\nu}\label{eq:TvDM}\\
&-\left(\zeta-\frac{2}{3}\eta_s \right) \nabla^\alpha u_{\dm\,\alpha}\left(\delta^\mu_\nu +u^\mu_\dm u_{\dm\,\nu} \right)\nonumber \\
&-\eta_s\Big(\nabla^\mu u_{\rm dm\, \nu}+\nabla_\nu u_{\rm dm}^{\mu}
+u_\dm^\alpha\nabla_\alpha(u^\mu_\dm u_{\dm\,\nu})\Big)\, . \nonumber
\end{align}
Here $u_\dm^\mu$ denotes the four-velocity of the dark matter and $\rho_\dm$ is the dark matter energy density. We consider models with negligible pressure such that $P_{\rm dm}=0$. The viscosity is encoded into two parameters, that we assume to be constant: $\zeta$ encodes the bulk viscosity, while $\eta_s$ accounts for the shear viscosity. Following~\cite{Zheng:2025oiq} we define two dimensionless parameters, related to $\zeta$ and $\eta_s$, that are convenient to write the equations 
\begin{align}
\ep=\frac{4\HH\eta_s}{a\bar{\rho}_\dm}   \quad\mbox{and}\quad  \Z=\frac{3\HH\zeta}{a\bar{\rho}_\dm} \, . \label{eq:vDM_epzeta}
\end{align}

The linear perturbation equations are derived in Appendix~\ref{app:vDM}. We see that the Poisson, gravitational slip and $0i$ Einstein equations are not modified. The continuity equations for baryons and dark matter are also standard, but the dark matter Euler equation is modified by the viscosity, as first studied in~\cite{Zheng:2025oiq}.

\subsubsection{Evolving dark energy}

We consider a scalar field dark energy model with an action 
\begin{align}
S^{\rm eDE}=\int d^4 x& \sqrt{-g}\Bigg[\frac{R}{16\pi G}-\frac{1}{2}\nabla_\mu\varphi\nabla^\mu\varphi\label{eq:SeDE}\\
&-U(\varphi)+\mathcal{L}_{\rm m}(\psi_{\rm sm}, \psi_{\rm dm}, g_{\mu\nu})\Bigg] \, . \nonumber 
\end{align}
Contrary to GBD theories and CQ models, here the scalar field is minimally coupled to matter (both baryonic and dark matter) and does not modify the form of the Einstein and conservation equations. The potential $U(\varphi)$ directly impacts the evolution of the scalar field and determines its equation of state $w_\de=\bar{P}_\de/\bar{\rho}_\de$, which depends on time and can differ from -1. The dark energy sound speed $c^2_{s\,\de}=\delta P_\de/\delta\rho_\de=1$ in such quintessence models and therefore the field remains essentially homogeneous inside the Hubble horizon.  

The linear perturbation equations are derived in Appendix~\ref{app:eDE}. We see that all equations keep the same form as in $\Lambda$CDM . 

\subsubsection{Clustering dark energy}

In clustering dark energy models, the scalar field exhibits non-negligible fluctuations inside the Hubble horizon, unlike in the quintessence model discussed above. One way to obtain significant dark energy clustering on observable scales is to have a very small sound speed~\cite{Creminelli:2009mu,Sefusatti:2011cm}. This can for example be obtained through non-standard kinetic terms in the action
\begin{align}
S^{\rm cDE}=\int d^4 x &\sqrt{-g}\Bigg[\frac{R}{16\pi G}+P(X)\label{eq:ScDE}\\
&+\mathcal{L}_{\rm m}(\psi_{\rm sm}, \psi_{\rm dm}, g_{\mu\nu})\Bigg] \, , \nonumber 
\end{align}
where $P(X)$ is a generic function of $X\equiv -\nabla^\mu\varphi\nabla_\mu\varphi$. The energy-density and pressure of dark energy are given by $\rho_\de=2P_{,X}X-P(X)$ and $P_\de=P(X)$. The form of the non-canonical kinetic terms, $P(X)$, determines the dark energy equation of state $w_\de=\bar{P}_\de/\bar{\rho}_\de$ and sound speed $c^2_{s\,\de}=\delta P_\de/\delta\rho_\de$. Following~\cite{Sefusatti:2011cm} we concentrate on models with a generic equation of state and with negligible sound speed $c^2_{s\,\de} \simeq 0$ which leads to dark energy fluctuations.\footnote{Note that some numerical simulations of clustering dark energy have found instabilities in the limit of very small sound speed~\cite{Hassani:2021tdd,Hassani:2022xyb}, while other codes do not exhibit such behavior~\cite{Blot:2026var}. The reason for such differences is still unclear, but if these models turn-out to be unstable, one can always choose a small but non-zero sound speed $c^2_s\gtrsim 10^{-4.7} $, for which instabilities are not present but dark energy is still clustering~\cite{Hassani:2022xyb}.}

The linear perturbation equations can be found in Appendix~\ref{app:cDE}. We see that the Poisson and gravitational slip equations are not modified. The $0i$ Einstein equation is affected by the flow of the scalar field. The continuity equation for dark energy differs from that of matter due to their different equation of state $w_\de\neq w_\m=0$. The Euler equations, however, are the same for matter and dark energy.

\subsection{Degeneracies}

Apart from viscous dark matter, all models rely on the presence of an additional degree of freedom. The difference between the three classes is then that: in dark energy models the new field is minimally coupled to gravity; in modified gravity the field is non-minimally coupled to gravity, influencing all matter components through a universal interaction; and in interacting dark matter models the field couples exclusively to dark matter, thereby breaking the weak equivalence principle for dark matter. Due to these differences, the Einstein and conservation equations take distinct forms in the five models. Therefore, if we could separately measure all the perturbed fields entering these equations, namely, the density fluctuations and peculiar velocities of baryons, dark matter, and dark energy, together with the two gravitational potentials $\Phi$ and $\Psi$, we could unambiguously distinguish among the five models. In practice, however, current surveys cannot measure these fields separately and are sensitive only to two combinations of them. 

First, spectroscopic galaxy surveys measure the galaxy correlation function, which depends on the galaxy density fluctuations $\delta_\g$ and galaxy velocities $V_\g$. The different multipoles of the correlation function allow us to measure the transfer functions of these two fields separately, see e.g.~\cite{2012PhRvD..86j3518P,eBOSS:2020yzd}.\footnote{The transfer function is the classical function $T_f(k,z)$ which relates a statistical field $f(\mathbf{k},z)$ to the primordial curvature perturbation $\mathcal{R}(\mathbf{k})$ generated by inflation: $f(\mathbf{k},z)=T_f(k,z)\mathcal{R}(\mathbf{k})$.} Second, cross-correlating cosmic shear or CMB lensing with galaxy clustering allows us to measure the transfer functions of $\delta_\g$ and of the Weyl potential $\Psi_W$ independently~\cite{Tutusaus:2023aux}.\footnote{CMB is also sensitive to the Integrated Sachs-Wolfe effect. This effect contains, however, the same information as lensing, since it depends on $\Psi_W$.} Among these three transfer functions measurable with current surveys, only those of $V_\g$ and $\Psi_W$ are relevant for testing the different models. Indeed, the galaxy density fluctuation $\delta_\g$ depends on the unknown galaxy bias, which is fully degenerate with the underlying density fluctuation $\delta$. 

The fact that we measure only the transfer functions of the fields, through their correlations, rather than the fields themselves is immaterial, since all the relevant equations can be recast in terms of transfer functions by dividing both sides by the primordial gravitational potential. In the following, to simplify the notation, we use the same symbols for the fields and their corresponding transfer functions; for example, $V_\g$ denotes the transfer function of the galaxy velocity. All tests are then formulated in terms of transfer functions. In practice, reconstructing the transfer functions as a function of redshift $z$ and scale $k$ may be complex and subject to uncertainties and systematics. For example, the relation between the shear-galaxy clustering cross-correlation and the transfer function of the Weyl potential depends on the comoving distances to the source and to the lens, which need to be measured from BAO and luminosity distances and reconstructed together with the transfer function. In this paper, we assume that this can be done with the desired accuracy, and we explore fundamental degeneracies between classes in the best case scenario.    

\begin{table}[th]
\renewcommand{\arraystretch}{1.5}
\centering
\caption{Explicit form of the four functions $\alpha,\Sigma,\sigma$ and $\gamma$ introduced in Eqs.~\eqref{eq:contgen}-\eqref{eq:evolgen}. These functions may depend on both time and scale. Some of them contain ratios of perturbations, which arise simply from the way Eqs.~\eqref{eq:contgen}-\eqref{eq:evolgen} are written, to highlight the degeneracies. These parameters can nonetheless be computed from the systems of equations in Appendix~\ref{app:equations}. In CQ and viscous dark matter the quantity $x$ denotes the fraction of dark matter $x=\bar{\rho}_\dm/\bar{\rho}_\m$. In clustering dark energy $x_\de$ is the fraction of dark energy $x_\de=\bar{\rho}_\de/\bar{\rho}_\m$.} \label{tab:params}
\begin{tabular}{c@{\hspace{1cm}} c}
\toprule
\multicolumn{2}{l}{\textbf{Generalized Brans-Dicke theories (GBD) }} \\
\midrule
$\alpha$ & 1   \\
$\Sigma$ & $A^2(\varphi)$\\
$\sigma$ & 1\\
$\gamma$ & $A^2(\varphi)\left(1+\frac{\beta^2}{3\beta^2+4\pi GA^2}\right)$ \\
\midrule
\multicolumn{2}{l}{\textbf{Coupled Quintessence (CQ) }}  \\
\midrule
$\alpha$ & 1   \\
$\Sigma$ & 1\\
$\sigma$ & $1+\beta x\frac{\bar{\varphi}'}{\HH}\frac{\delta'_\dm}{\delta'}$\\
$\gamma$ & $1+\frac{\beta^2 x^2}{4\pi G}\frac{\delta_\dm}{\delta}$ \\
\midrule
\multicolumn{2}{l}{\textbf{Viscous dark matter (vDM) }}  \\
\midrule
$\alpha$ & 1   \\
$\Sigma$ & 1\\
$\sigma$ & $1+x\frac{\ep+\Z}{3(1-\Z)}\left(\frac{k}{\HH}\right)^2\frac{\delta'_\dm}{\delta'}$\\
$\gamma$ & $1$ \\
\midrule
\multicolumn{2}{l}{\textbf{Evolving dark energy (eDE)}}  \\
\midrule
$\alpha$ & 1   \\
$\Sigma$ & 1\\
$\sigma$ & 1\\
$\gamma$ & 1\\
\multicolumn{2}{l}{\textbf{Clustering dark energy (cDE) }}  \\
\midrule
$\alpha$ & $\frac{\delta'_\m}{\delta'}$   \\
$\Sigma$ & 1\\
$\sigma$ & $\begin{aligned}
&\left[1+x_\de (1+w_\de) -\frac{x'_\de w_\de+x_\de w'_\de}{\HH} \right]\frac{\delta'_\m}{\delta'}\\& -\left[3 x_\de w_\de +\frac{x'_\de}{\HH} \right]\frac{\delta'_\de}{\delta'}\end{aligned}$\\
$\gamma$ & $\begin{aligned}1+x_\de(1+w_\de)+
\frac{(x'_\de+3\HH x_\de w_\de)'}{4\pi G a^2\bar{\rho}_\m}\frac{\delta_\de}{\delta}\end{aligned}$ \\
\bottomrule
\end{tabular} 
\end{table} 

To constrain the five models from the observation of the transfer functions, we combine the equations presented in Appendix~\ref{app:equations} into a system of three equations: a continuity equation for $V_\g$, a Poisson equation for $\Psi_W$, and a second-order evolution equation for the density fluctuation $\delta$. We find that this system takes the same general form in all five models:
\begin{align}
&kV_\g=\alpha\delta'\, ,\label{eq:contgen}\\
&\Psi_W=-4\pi G \frac{a^2}{k^2}\bar{\rho}_\m\Sigma\delta\, , \label{eq:Weylgen}\\
&\delta''+\HH\sigma\delta'-4\pi G a^2\bar{\rho}_\m\gamma\delta=0\, , \label{eq:evolgen}
\end{align}
where the functions $\alpha, \Sigma, \sigma$ and $\gamma$ depend on the model and all reduce to 1 in $\Lambda$CDM. Their explicit forms are given in Table~\ref{tab:params}. Note that the standard $\mu-\Sigma$ parametrisation of modified gravity~\cite{Zhao:2009fn,2011PhRvD..84h3523S}, would corresponds to $\alpha=\sigma=1$, $\gamma=\mu\neq 1$ and $\Sigma\neq 1$.

The procedure is then straightforward. For a given  model, we solve the evolution equation~\eqref{eq:evolgen}  and we compare the theoretical solution $\delta$ with the measured $V_\g$ and $\Psi_W$ using Eqs.~\eqref{eq:contgen} and~\eqref{eq:Weylgen}.\footnote{Of course in practice one does not necessarily use directly Eqs.~\eqref{eq:contgen}-\eqref{eq:evolgen}. One can instead solve the system of coupled equations presented in Appendix~\ref{app:equations} and compare the theoretical predictions for $V_\g$ and $\Psi_W$ with the measurements. However, independently of the way the analysis is performed, the information on a given model does come from the combination of Eqs.~\eqref{eq:contgen}-\eqref{eq:evolgen}.} This leads to two conclusions:

\vspace{0.1cm}

\noindent\textbf{Conclusion 1}: Within a given model, we see from the expressions for $\alpha, \Sigma, \sigma$ and $\gamma$ in Table~\ref{tab:params} that there is typically enough information to constrain several parameters of the model. In particular, combining the measurement of $V_\g$ and $\Psi_W$ is more powerful than using either observable alone, since the functions $\alpha, \Sigma, \sigma$ and $\gamma$ depend differently on the model parameters. For example, in GBD theories, one can constrain both $A(\varphi)$ and its derivative $\beta$ from the combination of $V_\g$ and $\Psi_W$. Similarly, in the standard $\mu-\Sigma$ parametrization of modified gravity, $\Psi_W$ is sensitive to a combination of $\Sigma$ and $\mu$, while $V_{\rm g}$ constrains $\mu$ (since $\alpha=\gamma=1$) allowing to separate these two functions~\cite{2011PhRvD..84h3523S}.

\vspace{0.1cm}

\noindent\textbf{Conclusion 2}: This method is unable to discriminate among the different models. In all five models, the evolution of $\delta$ differs from that predicted by $\Lambda$CDM. Even in the case of evolving dark energy, where $\alpha= \Sigma=\sigma=\gamma=1$, the evolution is modified through $\HH$, which differs from its $\Lambda$CDM value. As a consequence, any deviation of the measured $V_\g$ and/or $\Psi_W$ from the $\Lambda$CDM predictions can be reproduced by more than one model. The fact that clustering dark energy is the only model with $\alpha\neq 1$ (continuity equation modified) and that GBD theories are the only ones with $\Sigma\neq 1$ (Poisson equation modified) does not help in breaking degeneracies, since $\delta$ is not measured. We always compare a theoretical prediction for $\delta$ with the measured $V_\g$ and $\Psi_W$, and consequently any $\Sigma\neq 1$ or $\alpha\neq 1$ can be mimicked by a change in the theoretical $\delta$ (i.e.\ $\sigma\neq 1$ and/or $\gamma\neq 1$). This conclusion also applies to the standard $\mu-\Sigma$ parametrization of gravity, which is degenerate with dark matter or dark energy models with non-trivial $\sigma$ and/or $\alpha$.
The only model which could, in some circumstances, be differentiated from the others is evolving dark energy. In this model the expansion rate of the Universe and the growth of structure are linked since they are both modified by the same quantity $\bar{\rho}_{\de}$ (which modifies $\HH$). As such, the measurement of the background expansion rate (from distances), puts constraints on the possible deviations in $V_\g$ and $\Psi_W$. This could in principle be mimicked by cancellations between the functions $\alpha, \Sigma, \sigma$ and $\gamma$, but it would be a fine-tuned solution.

This degeneracy among the different models can be understood intuitively. The depth of the Weyl potential $\Psi_W$ can indeed be modified by changing the expansion rate of the Universe (evolving dark energy); by modifying the link between the metric and the density fluctuations (modified gravity); by introducing another fluid which clusters and distorts the  metric (clustering dark energy); or by modifying the growth of dark matter perturbations through additional forces or viscosity (extended dark matter). Similarly galaxy velocities are affected by the depth of the potentials, by additional forces and by viscosity. Hence, all models can generate similar differences in $\Psi_W$ and in $V_\g$ with respect to the $\Lambda$CDM predictions.

To overcome this degeneracy, the way forward is to compare measurable quantities directly with one another. Rather than comparing theoretical predictions derived within a given model or class of models with measurements of $\Psi_W$ and $V_\g$, one can directly test fundamental equations that take different forms in the five models.

\section{Level one: Combining the Weyl potential and galaxy velocities to discriminate between models}
\label{sec:level1}

With this idea in mind, the gravitational slip equation has been proposed as a way to discriminate among models by comparing the temporal and spatial distortions of the metric~\cite{Song:2010rm,EuclidTheoryWorkingGroup:2012gxx,Motta:2013cwa,Amendola:2016saw}. In GBD theories, the two gravitational potentials differ, as is generically the case in modified gravity models, whereas dark energy and extended dark matter models satisfy $\Psi=\Phi$. A measurement of $\Psi\neq\Phi$ would therefore constitute a smoking gun for modified gravity. However, as explained in~\cite{Bonvin:2022tii}, this test cannot be implemented with current data, since only $\Psi_W$ and $V_\g$ are measured. Here we do not repeat the argument developed in~\cite{Bonvin:2022tii}, but we simply rephrase it in terms of $V_\g$ and $\Psi_W$ for the five models considered. 

In GBD theories we have that
\begin{align}
\Phi-\Psi=-2\beta\delta\varphi\, .  \label{eq:anis} 
\end{align}
Eq.~\eqref{eq:anis} can be written in terms of $V_{\rm g}$ and $\Psi_W$ using that 
\begin{align}
\Phi=2\Psi_W-\Psi\, ,  \label{eq:PhiPsiW} 
\end{align} 
and relating $\Psi$ to $V_{\rm g}$ through Euler equation~\eqref{eq:EulergGBD}
\begin{align}
\Psi=-\frac{1}{k}(V'_\g+\HH V_\g)\, .  \label{eq:PsiV}  
\end{align}
Using Eqs.~\eqref{eq:PhiPsiW} and~\eqref{eq:PsiV} we can build a test $\teststress$ encoding the difference between the two gravitational potentials, written in terms of observable quantities only
\begin{align}
\teststress\equiv \Psi_W+\frac{1}{k}(V'_\g+\HH V_\g)\, .\label{eq:deftest}  
\end{align}
In GBD theories we have
\begin{align}
\teststress_{\rm GBD}=\frac{\Phi-\Psi}{2}=-\beta\delta\varphi\neq 0\, . 
\end{align}
Measuring $\teststress\neq 0$ would therefore indicate that gravity is modified. 

However, computing $\teststress$ for the four other models using Eq.~\eqref{eq:deftest} we find:
\begin{align}
\teststress_{\rm CQ}&=-\frac{ \beta x}{k}\left[\varphi' V_{\rm dm}+k\delta\varphi\right]\neq 0\, , \\
\teststress_{\rm vDM}&=- x\frac{\ep+\Z}{3(1-\Z)}\frac{k}{\HH}V_\dm\neq 0\, , \\
\teststress_{\rm eDE}&=0\, ,\\
\teststress_{\rm cDE}&=0\, ,
\end{align}
where $x=\bar{\rho}_\dm/\bar{\rho}_\m$. A measurement consistent with $\teststress=0$ would therefore be compatible with $\Lambda$CDM, evolving dark energy, and clustering dark energy, whereas a non-zero value would be compatible with GBD theories, CQ models, and viscous dark matter. More generally, $\teststress\neq0$ arises both in models that modify the relation between $\Phi$ and $\Psi$ and in models that modify the Euler equation, i.e.\ the relation between $V_\g$ and $\Psi$. This test is therefore unable to discriminate among the different models and, in particular, does not provide a smoking gun for modified gravity.

\section{Level two: The three master equations that break them all apart}
\label{sec:level2}

To break the degeneracies among the different models, two novel ingredients are required. First, we need additional information, since Section~\ref{sec:level1} showed that $V_\g$ and $\Psi_W$ alone are insufficient. Second, we need a set of tests that rely only on observable quantities. Comparing observables with theoretical predictions, as done in Section~\ref{sec:level0}, does not work because the relevant equations depend on the class of models under consideration, which is not known a priori. Restricting the analysis to a specific class by fixing the equations also fails, since this artificially ignores degeneracies with other classes and can lead to erroneous conclusions. Finally, replacing an unobservable quantity with an observable one, as done in Section~\ref{sec:level1}, is not a viable solution either, since it relies on specific equations that are violated in some classes of models.

\subsection{A set of three tests}

Gravitational redshift provides the missing observational ingredient. It has already been measured in galaxy clusters~\cite{Wojtak:2011ia,Sadeh:2014rya, Jimeno:2014xma, eBOSS:2021ofn, Rosselli:2022qoz} and will become accessible on cosmological scales in the coming years~\cite{Breton:2018wzk,Beutler:2020evf,Saga:2021jrh,Sobral-Blanco:2022oel, Bonvin:2023jjq}. Unlike gravitational lensing, this effect is sensitive solely to the temporal distortion of the metric and provides a direct measurement of $\Psi$. The goal is then to combine measurements of $\Psi$, $\Psi_W$, and $V_\g$ in a way that breaks the degeneracies among the five models. 

\begin{table*}[th]
\renewcommand{\arraystretch}{1.4}
\centering
\caption{Outcomes of the three tests defined in Eqs.~\eqref{eq:Nslip}-\eqref{eq:Neuler} for the five models considered. The last column summarizes the result of each test, with 0 indicating that the test vanishes and X that it is non-zero. The five models lead to distinct combinations of outcomes, allowing them to be distinguished unambiguously.} \label{tab:tests}
\begin{tabular}{c@{\hspace{1cm}}c@{\hspace{1cm}}c}
\toprule
\multicolumn{3}{l}{\textbf{GBD} } \\
\midrule
$\testij$ & $-\beta\delta\varphi$  & X  \\
$\testzi$ & $\beta\bar{\varphi}'\Psi+4\pi G a^2\bar{\rho}_\m(1-A^2)\frac{V_\g}{k}-\beta\delta\varphi'+\left(4\pi G A^2\bar{\varphi}'+\beta\HH+2\beta^2\bar{\varphi}'-\beta_{,\varphi}\bar{\varphi}'\right)\delta\varphi$ & X\\
$\testeuler$ & 0 & 0\\
\midrule
\multicolumn{3}{l}{\textbf{CQ} } \\
\midrule
$\testij$ & 0 & 0   \\
$\testzi$ & $4\pi G \bar{\varphi}'\delta\varphi$ & X\\
$\testeuler$ & $-\beta x\left(\bar{\varphi}'V_\dm+k\delta\varphi \right)$ & X \\
\midrule
\multicolumn{3}{l}{\textbf{Viscous dark matter} }\\
\midrule
$\testij$ & 0 & 0   \\
$\testzi$ & $0$ & 0\\
$\testeuler$ & $-x\HH\frac{\ep+\Z}{3(1-\Z)}\left(\frac{k}{\HH} \right)^2V_\dm$ & X \\
\midrule
\multicolumn{3}{l}{\textbf{Evolving dark energy} } \\
\midrule
$\testij$ & 0 & 0   \\
$\testzi$ & 0 & 0\\
$\testeuler$ & 0& 0\\
\midrule
\multicolumn{3}{l}{\textbf{Clustering dark energy}  } \\
\midrule
$\testij$ & 0 & 0   \\
$\testzi$ & $4\pi G a^2\bar{\rho}_\de(1+w_\de)\frac{\delta\varphi}{\bar{\varphi}'}$ & X\\
$\testeuler$ & 0& 0\\
\bottomrule
\end{tabular} 
\end{table*}

Since these quantities enter three master equations -- the gravitational slip equation, the $0i$ Einstein equation, and the Euler equation -- we construct three tests directly probing these equations. Strictly speaking, the gravitational slip equation and the $0i$ Einstein equation depend on $\Phi$, but from the definition of $\Psi_W$ we can write $\Phi=2\Psi_W-\Psi$. The tests are then given by: 
\begin{align}
\testij&\equiv \Psi_W-\Psi\, , \label{eq:Nslip}\\
\testzi&\equiv 2\Psi'_W-\Psi'+\HH\Psi+4\pi G a^2\bar{\rho}_\m\frac{V_\g}{k}\, ,\label{eq:N0i}\\
\testeuler&\equiv V'_\g+\HH V_\g+k\Psi\, . \label{eq:Neuler}
\end{align}
These three tests vanish identically in $\Lambda$CDM. They can be directly measured by combining observations of $\Psi$, $\Psi_W$, and $V_\g$,\footnote{The tests also depend on derivatives of these quantities, which can be inferred from measurements at different redshifts, either directly through numerical derivatives constructed from neighboring redshift bins or through interpolation and reconstruction methods such as splines or Gaussian processes.} together with measurements of $\HH$ and $\bar{\rho}_\m$ obtained from distances. Their outcomes then provide a diagnostic for correctly identifying the underlying model. Table~\ref{tab:tests} summarizes the values of the three tests for the five models. As shown in the last column, the five models lead to distinct outcomes and can therefore be distinguished unambiguously. Evolving dark energy leaves all three tests at zero, but can still be distinguished from $\Lambda$CDM through measurements of the expansion history.

From Table~\ref{tab:tests}, we see that the combination of all three tests is necessary to fully break the degeneracies. The test $\testij$ provides a key diagnostic of modified gravity. Scalar field dark energy models satisfy indeed $\testij=0$. Even when dark energy clusters and contributes to the perturbations, it does not possess an anisotropic stress and hence does not generate a difference between the two gravitational potentials. Extended dark matter models likewise satisfy $\testij=0$. Viscous dark matter can, in principle, generate a difference between the two gravitational potentials. However, as shown in Eq.~\eqref{eq:ijvdm} and discussed in~\cite{Zheng:2025oiq}, this difference is suppressed by a factor $(\HH/k)^2$ relative to the effect on the Euler equation and is therefore negligible on observable scales.

The test $\testeuler$ is an essential ingredient for identifying deviations from geodesic motion. Such deviations can arise either from a violation of the weak equivalence principle in the dark matter sector, due to additional forces acting on dark matter, or from microscopic processes that generate viscosity and thereby modify dark matter motion. A non-zero $\testeuler$ therefore singles out extended dark matter and cannot be mimicked by modified gravity or dark energy.

Finally, the test $\testzi$ complements the other two tests by identifying the presence of additional clustering degrees of freedom. It can therefore distinguish between a smooth dark energy component ($\testzi=0$) and a clustering dark energy component ($\testzi\neq 0$). Similarly, it can discriminate between an additional force acting on dark matter ($\testzi\neq 0$) and microscopic processes modifying the dark matter energy–momentum tensor ($\testzi=0$)~\footnote{Ref.~\cite{Zheng:2025oiq} argues that a breaking of the weak equivalence principle for dark matter can be differentiated from viscosity using the Euler equation alone, by exploiting the different scale-dependence of these two effects. However, in full generality a breaking of the weak equivalence principle can also depend on scale (for example, if it arises from interactions of dark matter with dark radiation or from self-interactions through collisions) which would make it indistinguishable from viscosity if $\testzi$ is not used.}. Together, the three tests can therefore determine unambiguously the nature of deviations from $\Lambda$CDM.

\subsection{Generalization}

Here we have considered five representative models spanning the three classes, showing that the three tests can distinguish between them. The landscape of models beyond $\Lambda$CDM is, however, much broader, which raises the question of how general our conclusions are and for which models the tests may remain degenerated.

First, although generic extensions of cold dark matter do not have an intrinsic anisotropic stress, it is possible to construct phenomenological dark matter models with such features, see e.g.~\cite{Hu:1998kj}. However, as shown in Appendix~\ref{app:vDM} such an anisotropic stress would generate corrections to the Euler equation that are enhanced by a factor $(k/\HH)^2$ with respect to corrections in the gravitational slip equation. The reason is that the Euler equation is affected by the force generated by the gradient of the anisotropic stress, whereas the gravitational slip equation depends directly on the anisotropic stress itself. Hence, any non-negligible deviation in the gravitational slip equation would lead to very large modifications to the Euler equation, that would already have been ruled out. In practice these models are therefore characterized by $\testeuler\neq 0$ and $\testzi=\testij\simeq 0$. From Table~\ref{tab:tests} we see that such models would be correctly identified as extended dark matter models with departures from the perfect-fluid behavior. They are therefore not degenerated with dark energy and modified gravity models.

Second, dark matter may experience a fifth force mediated by a field whose background energy density is subdominant and therefore does not contribute appreciably to the energy budget of the Universe. In this case, we would have $\testeuler\neq0$ while $\testzi=0$. Such models would still be correctly identified as extended dark matter and could not be misinterpreted as modified gravity or dark energy. However, Table~\ref{tab:tests} shows that they would remain degenerate with viscous dark matter. 

Third, while most dark matter interactions involve momentum exchange, it is possible to construct specific models where dark matter exchanges only energy but not momentum with other fields or particles. In this case, the continuity equation is modified, whereas the Euler equation remains unchanged. Since our framework does not test the continuity equation, such models may not be correctly identified. Depending on the model, the component receiving or mediating the energy exchange may or may not contribute to the total momentum density, so that $\testzi$ may be either zero or non-zero. As shown in Table~\ref{tab:tests}, these models would therefore be degenerated with either evolving dark energy or clustering dark energy. Testing the continuity equation would break the degeneracy between interacting dark matter and evolving dark energy. However, interacting dark matter could remain degenerated with clustering dark energy, since the continuity equation is modified in both cases. The form of the modifications may differ sufficiently to distinguish the two scenarios, but this remains to be investigated. Moreover, testing the continuity equation would require new methods to measure the density fluctuation $\delta$, for example by combining the galaxy power spectrum and bispectrum to break the degeneracy between $\delta$ and galaxy bias~\cite{2000ApJ...544..597S,2002astro.ph.12311V,2006PhRvD..74b3522S,Gil-Marin:2014sta,2021MNRAS.501.2862S}.

Fourth, modified gravity could be degenerated with clustering dark energy for models that do not generate $\Psi\neq\Phi$. Although broad classes of modified gravity models alter the relation between $\Phi$ and $\Psi$, specific models satisfying $\Psi=\Phi$ can be constructed~\cite{Linder:2018jil}. Such models would have only $\testzi$ that is non-zero, which is the same as the outcome of clustering dark energy. This degeneracy could in principle be broken by extending the analysis to smaller scales, such as cluster scales or below. In viable modified gravity models, deviations from General Relativity  need to be screened at small scales to be compatible with Solar System tests, driving $\testzi$ towards zero. The clustering of dark energy, on the other hand, could generically persist at small scales maintaining $\testzi\neq0$.

Finally, evolving dark energy could also lead, in some specific cases, to $\testzi\neq 0$. As first pointed out in~\cite{Amendola:2012ky}, if part of the dark energy component evolves as $1/a^3$, it would be fully degenerated with matter. As a consequence the $\Omega_{\m 0}$ that is inferred from the evolution of $\HH(z)$ (through measurements of the distance-redshift relations) would be incorrect since it would contain part of the dark energy density. This could lead to $\testzi\neq 0$ in Eq.~\eqref{eq:N0i} due to the fact that $\bar{\rho}_\m$ is not the true background matter density. From Table~\ref{tab:tests} we see that such an evolving dark energy model would still be correctly identified as dark energy and would not be confused with modified gravity or extended dark matter. However, it could mimic clustering dark energy.   

From this we conclude that our results are much more general than the five examples we considered and hold for a vast range of models within the three classes. This is not surprising since the tests are targeted at fundamental equations that are representative of the three classes. It is however possible to construct specific models that remain degenerated under the results of the three tests. Most of these models are degenerated within their own class: dark matter interactions compared to beyond perfect fluid dark matter, or evolving dark energy compared with clustering dark energy. Two specific models can however mimic models of another class, namely dark matter models with a pure energy exchange and modified gravity with $\Psi=\Phi$. Measuring $\delta$ and testing the continuity equation may solve the first degeneracy, while extending the tests to cluster scales could solve the second one.

\section{Information content: comparison of Einstein's equations}
\label{sec:Poisson}

In constructing our tests we have argued that the $0i$ Einstein equation should replace Poisson equation, which cannot be tested directly since $\delta$ is not measurable. Current tests of gravity, which assume standard cold dark matter, have instead been designed to constrain the two phenomenological functions $\mu$ and $\Sigma$, which modify the Poisson equations for $\Psi$ and $\Psi_W$:~\footnote{The Poisson equation~\eqref{eq:PoissonPsiW} for $\Psi_W$ is a combination of the Poisson equation for $\Psi$ and the gravitational slip equation.}
\begin{align}
k^2\Psi&= -4\pi G\bar{\rho}_\m a^2\mu\delta\, ,\\
k^2\Psi_W&= -4\pi G\bar{\rho}_\m a^2\Sigma\delta\, .\label{eq:PoissonPsiW}
\end{align}
One may then wonder whether the two tests $\testij$ and $\testzi$ provide the same information as the two Poisson equations when we restrict ourselves to the class of modified gravity models. To show that this is indeed the case and that no information is lost, we compare the two approaches in GBD theories.

We use the Klein–Gordon equation~\eqref{eq:KGGBD} to rewrite $\delta\varphi$ and $\delta\varphi'$ appearing in $\testzi$ in Table~\ref{tab:tests} in terms of $\Psi$ and $\Psi_W$. We obtain
\begin{align}
\testzi_{\rm GBD}=&4\HH\left[2\tilde{\beta}^2 +\nu +4\nu(2\tilde{\beta}^2-\epsilon)\right]\Psi_W \label{eq:ziGBD}\\
&-2\HH\left[3\tilde{\beta}^2+\nu+6\nu(2\tilde{\beta}^2-\epsilon) \right]\Psi \nonumber\\
&-8\tilde{\beta}^2\Psi'_W+6\tilde{\beta}^2\Psi'+(1-A^2)4\pi G\bar{\rho_\m}\frac{V_\g}{k}\, , \nonumber
\end{align}
where
\begin{align}
\tilde{\beta}^2&=\frac{\beta^2}{8\pi G A^2}\, , \\
\epsilon&=\frac{\beta_{,\varphi}}{8\pi G A^2}\, ,\\
\nu&=\frac{\beta\bar{\varphi}'}{\HH}\,. \label{eq:nu}
\end{align}
Similarly, $\testij$ can be written as 
\begin{align}
 \testij_{\rm GBD}=6\tilde{\beta}^2\Psi-8\tilde{\beta}^2\Psi_W\, .  
\end{align}
Combined, the two tests are therefore sensitive to four quantities: $\tilde{\beta}^2$, $A^2-1$, $\nu$ and $\epsilon$. In practice, the effect of $\epsilon$ is generally strongly suppressed because it always appears multiplied by $\tilde{\beta}^2$. 

We can now compare this with the $\mu-\Sigma$ approach. From Eqs.~\eqref{eq:PoissonGBD} and~\eqref{eq:WeylGBD} we obtain
\begin{align}
\mu&=A^2\left(1+\frac{2\tilde\beta^2}{6\tilde\beta^2+1}\right)\, ,\\
\Sigma&=A^2\, ,
\end{align}
which shows that $\mu$ and $\Sigma$ constrain two quantities: $\tilde{\beta}^2$ and $A^2-1$. 

We see therefore that no information is lost when replacing the two Poisson equations by $\testij$ and $\testzi$. The two dominant deviations from General Relativity generated by GBD theories, $\tilde{\beta}^2$ and $A^2-1$, are indeed captured by $\testij$ and $\testzi$. Interestingly, $\testzi$ is also sensitive to $\nu$, which does not contribute to the two Poisson equations. From Eq.~\eqref{eq:nu}, we see that $\nu$ depends on the time derivative of $\bar{\varphi}$. In models where the scalar field remains close to its minimum and varies little throughout cosmic evolution, $\nu$ is negligible compared to $\tilde{\beta}^2$ and $A^2-1$. However, in models where the scalar field evolves on a timescale comparable to the Hubble time, $\nu$ becomes relevant, generating additional modifications to $\testzi$, that are not directly captured by the $\mu-\Sigma$ approach. This missing information can however in principle be reconstructed from the redshift evolution of $A^2-1$, which depends on $\bar{\varphi}'$.

The difference in sensitivity between the $\mu-\Sigma$ approach and the tests $\testij$ and $\testzi$ comes from the fact that, in the two Poisson equations, time derivatives of the scalar field, $\bar{\varphi}'$, impact the equations beyond the Quasi-Static Approximation (QSA). These modifications are therefore strongly suppressed inside the horizon and neglected in the $\mu-\Sigma$ approach. $\testzi$, on the other hand, has the same form with and without the QSA, since all terms contain the same number of gradients. As a consequence  $\tilde\beta^2$ and $A^2-1$ impact $\testzi$ at the same order in $k/\HH$ as $\nu$. The price to pay for this additional sensitivity of $\testzi$ is that $A^2-1$ is degenerated with $\nu$ in that test and can therefore not be reconstructed independently. 

In summary, this calculation shows that within the class of modified gravity models, no information is lost when using the three tests instead of the $\mu-\Sigma$ approach: any deviation detected by the $\mu-\Sigma$ approach would also be captured by the $\testij$ and $\testzi$ tests. In addition, the three tests have the advantage that they can discriminate modified gravity from extended dark matter and dark energy, which is not possible with the $\mu$–$\Sigma$ approach.

\section{Conclusion}
\label{sec:conclusion}

In this paper, we have proposed a set of three tests targeted at identifying deviations from the $\Lambda$CDM model. These tests are constructed from fundamental equations that are modified by the three key classes of models beyond $\Lambda$CDM: extended dark matter, dark energy and modifications of gravity. The essential feature of these tests is that they rely exclusively on measurable quantities: the galaxy velocity, the Weyl potential and the time distortion. As such no assumption needs to be made about the underlying theory of gravity, dark energy or dark matter to apply these tests.  We have demonstrated that the results of the tests provide a clean and robust way to discriminate between the three classes.

Although expressed differently, these tests extend the ones presented in~\cite{Amendola:2012ky}, which combine ratios of observables to detect departures from $\Lambda$CDM in a model-independent way.
As explained in that work, the motivation for using ratios of observables is to remain independent of the shape of the matter density fluctuations today, which is unknown if dark energy is allowed to cluster. The three tests constructed here automatically incorporate this feature by testing the relations between different observables, without assuming any specific shape for the matter density fluctuations. Our tests, however, go beyond the framework of~\cite{Amendola:2012ky} by including gravitational redshift, which allows a direct measurement of the distortion of time. This additional observable is essential for constructing $\testzi$ and $\testeuler$, which cannot be tested within the framework of~\cite{Amendola:2012ky}. In that approach, the Euler and continuity equations are assumed to hold, and $\Psi$ is effectively reconstructed from velocity measurements.

Compared with other model-independent tests in the literature, like the $E_G$ statistics~\citep{Zhang:2007nk,Alam:2016qcl,Amon2018MNRAS.479.3422A, Reyes:2010tr, Blake:2016fcm, Pullen2016, MoradinezhadDizgah:2016pqy, Ghosh:2018ijm, Wenzl:2024sug, Abidi_2023_10419,Grimm:2024fui}, the scale-invariance of the growth of structure~\citep{Franco:2019wbj} or the consistency between the growth of the Weyl potential and that of the density~\cite{Uzan:2000mz,Uzan:2006mf,Bonvin:2025qeo}, that are targeted at some specific feature of beyond $\Lambda$CDM models, the tests $\testij, \testzi$ and $\testeuler$ have as primary goal to discriminate between different classes of models. They therefore provide a first powerful diagnostic to identify which part of $\Lambda$CDM is modified, which can then be refined by applying more specific tests within the relevant class of models.

In general, our tests will be less constraining than standard analyses applied to a specific model. They rely indeed on measurements of gravitational redshift, which are less precise than velocity and lensing measurements. As such, within a given class of models, our tests cannot compete with standard methods. However, the power of the tests is to unambiguously identify the correct sector which is modified, if a deviation is found in the data. Without these tests, we may be in a situation where part of the community claims a breaking of General Relativity, while the other part announces new interactions in the dark sector. Only the combination of the three tests can identify, without mistake, which claim is the correct one. In a forthcoming paper, we will quantify the discriminating power of the tests for the upcoming generation of surveys. 

\section*{Acknowledgments}

We thank Ruth Durrer and Levon Pogosian for interesting discussions and valuable feedback on the draft of this paper. C.B. acknowledges support from the Swiss National Science Foundation. 

\newpage

\appendix

\onecolumngrid

\section{Equations}
\label{app:equations}

\subsection{Generalized Brans-Dicke theories (GBD)}
\label{app:GBD}

The action for GBD is given in Eq.~\eqref{eq:SGBD}. We work in the Quasi-Static Approximation (QSA) and we consider model with a negligible mass of the scalar field, i.e.\ such that
\begin{align}
m^2\equiv U_{,\varphi\varphi}+\beta_{,\varphi}\rho\ll \frac{k^2}{a^2}\quad \mbox{where}\quad \beta\equiv\frac{A_{,\varphi}}{A} \, ,  
\end{align}
for all scales $k$ of interest. This is the case, e.g., in models where the mass of the scalar field is lighter than or equal to the size of the Universe today $m \lesssim H_0$. This condition restricts the range of models considered but it has the advantage of simplifying the calculations without impacting the degeneracy argument. The linearized Einstein, conservation and Klein-Gordon equations for these models can be derived from the action and read:
\begin{align}
&{\rm Einstein\ 00\ (Poisson)}: k^2\Phi=-4\pi G a^2 A^2 \delta\rho-\beta k^2\delta\varphi\, , \label{eq:00GBD}\\   
&{\rm Einstein\ } i\neq j: \Phi-\Psi=-2\beta\delta\varphi\, ,\label{eq:ijGBD}\\
&{\rm Einstein\ } 0i: \Phi'+(\HH-\beta\bar{\varphi}')\Psi=-4\pi Ga^2A^2\bar{\rho}_\m\frac{V}{k}-\beta\delta\varphi'+(4\pi G A^2\bar{\varphi}'+\beta\HH+2\beta^2\bar{\varphi}'-\beta_{,\varphi}\bar{\varphi}')\delta\varphi\, ,\label{eq:0iGBD} \\
&{\rm Klein-Gordon}: \delta\varphi=\frac{\beta}{4\pi G A^2}(2\Phi-\Psi)\, ,\label{eq:KGGBD}\\
&{\rm Continuity}: \delta'=k V\, ,\label{eq:contGBD}\\
&{\rm Euler}: V'+\HH V+k\Psi=0\, .\label{eq:EulerGBD}
\end{align}
Here $\delta=\delta\rho/\bar{\rho}_\m=(\delta\rho_\dm+\delta\rho_\ba)/(\bar{\rho}_\dm+\bar{\rho}_\ba)$ is the total density fluctuation, composed of dark matter and baryons and $V$ is the total peculiar velocity which is equal to the individual velocity of the baryons and dark matter since both components obey the same Euler equation: $V=V_\dm=V_\ba$. The scalar field $\varphi$ is split into its background part $\bar\varphi$ and its perturbation $\delta\varphi$. For a given GBD model, i.e.\ for a given form of $A(\varphi)$, one can solve the coupled system of equations~\eqref{eq:00GBD} to~\eqref{eq:EulerGBD} to find the solutions for all the fields. Assuming that there is no velocity bias, which is a good approximation at large scales~\cite{Zheng:2014vla,Angulo:2021kes}, the galaxy velocity is then directly given by the matter velocity $V_\g=V$.

To demonstrate the degeneracy between the different models, it is useful to rewrite the system of equations in the following way. First, combining Eqs.~\eqref{eq:00GBD},~\eqref{eq:ijGBD} and~\eqref{eq:KGGBD} the scalar field perturbation can be related to the matter density through
\begin{align}
\delta\varphi=\frac{-\beta a^2\delta\rho}{k^2\left(1+\frac{3\beta^2}{4\pi G A^2}\right)}\, . \label{eq:varphideltaGBD}   
\end{align}
Inserting Eq.~\eqref{eq:varphideltaGBD} into Eqs.~\eqref{eq:00GBD} and~\eqref{eq:ijGBD} give rise to two modified Poisson equations:
\begin{align}
\Psi&=-4\pi G \frac{a^2}{k^2}A^2\bar{\rho}_\m\delta\left(1+\frac{\beta^2}{3\beta^2+4\pi GA^2}\right)\, , \label{eq:PoissonGBD}\\
\Psi_W&=-4\pi G \frac{a^2}{k^2}A^2\bar{\rho}_\m\delta\label{eq:WeylGBD}\, .
\end{align}
The galaxy velocity can be related to the matter density through the continuity equation~\eqref{eq:contGBD}
\begin{align}
k V_{\rm g} = \delta' \, ,\label{eq:contgGBD}
\end{align}
and it obeys Euler equation
\begin{align}
V_\g'+\HH V_\g+k\Psi=0\, .\label{eq:EulergGBD}    
\end{align}
Finally, combining the Euler~\eqref{eq:EulerGBD}, continuity~\eqref{eq:contGBD} and Poisson equations~\eqref{eq:PoissonGBD}, we can derive a second-order evolution equation for $\delta$
\begin{align}
\delta''+\HH\delta'-4\pi G a^2 \bar{\rho}_\m A^2  \left(1+\frac{\beta^2}{3\beta^2+4\pi GA^2}\right)\delta=0\, .  \label{eq:evolGBD}
\end{align}
For a given model, the evolution equation~\eqref{eq:evolGBD} can be solved, and the solution can then be compared with the measured $\Psi_W$ and $V_\g$ using Eqs.~\eqref{eq:WeylGBD} and~\eqref{eq:contgGBD}.

\subsection{Coupled Quintessence (CQ)}
\label{app:CQ}

The action for CQ is given in Eq.~\eqref{eq:SCQ}. As for GBD we work in the QSA and we consider models with a negligible mass such that the linearized equations take the form: 
\begin{align}
&{\rm Einstein\ 00\ (Poisson)}: k^2\Phi=-4\pi G a^2 \delta\rho\, ,  \label{eq:00CQ}\\   
&{\rm Einstein\ } i\neq j: \Phi-\Psi=0\, , \label{eq:ijCQ}\\
&{\rm Einstein\ } 0i: \Phi'+\HH\Psi=4\pi G\left(\bar{\varphi}'\delta\varphi-a^2\bar{\rho}_{\rm b}\frac{V_{\rm b}}{k}-a^2\bar{\rho}_{\rm dm}\frac{V_{\rm dm}}{k} \right)\, ,\\
&{\rm Klein-Gordon}: \delta\varphi=\frac{-\beta a^2\delta\rho_{\rm dm}}{k^2}\, , \\
&{\rm Continuity\ dm}: \delta'_{\rm dm}=k V_{\rm dm}\, , \label{eq:contCQdm}\\
&{\rm Continuity\ b}: \delta'_{\rm b}=k V_{\rm b}\, , \label{eq:contCQb}\\
&{\rm Euler\ dm}: V'_{\rm dm}+(\HH+\beta\bar{\varphi}') V_{\rm dm}+k\Psi=-k\beta\delta\varphi\, ,  \label{eq:EulerCQdm}\\
&{\rm Euler\ b}: V'_{\rm b}+\HH V_{\rm b}+k\Psi=0\, .\label{eq:EulerCQb}
\end{align}

The galaxy velocity is a weighted average of the baryon and dark matter density
\begin{align}
V_{\rm g}=V=x V_{\rm dm}+(1-x)V_{\rm b}\, ,   \label{eq:VgCQ} 
\end{align}
where $x$ denotes the fraction of dark matter with respect to the total matter $x\equiv \bar{\rho}_{\rm dm}/\bar{\rho}_\m$. If there were no interaction between dark matter and the scalar field, then $V_\ba=V_\dm$ and Eq.~\eqref{eq:VgCQ} would lead to $V_\g=V_\dm=V_\ba$ as in GBD. However, the interaction term in Eq.~\eqref{eq:SCQ} leads to additional contributions to the Euler equation for dark matter~\eqref{eq:EulerCQdm}, generating a difference between the velocities of baryons and dark matter. As shown in~\cite{Bonvin:2022tii}, the velocity measured from redshift-space distortion is that of the center of mass of the galaxies, $V_\g$ given in Eq.~\eqref{eq:VgCQ}. Combining Eqs.~\eqref{eq:EulerCQdm} and~\eqref{eq:EulerCQb} we find that this velocity obeys the modified Euler equation
\begin{align}
V_\g'+\HH V_\g+k\Psi=-x\beta\big(\bar{\varphi}'V_\dm+k\delta\varphi\big)\, , \label{eq:EulerCQ}
\end{align}
Assuming that $x$ remains approximately constant over time, the galaxy velocity can be related to the matter density using the continuity equations~\eqref{eq:contCQdm} and~\eqref{eq:contCQb} 
\begin{align}
kV_{\rm g}&=\delta'\, . \label{eq:contgCQ}\\
\end{align}
Since $\Phi=\Psi$, the Poisson equations for $\Phi, \Psi$ and $\Psi_W$ are the same
\begin{align}
\Phi=\Psi=\Psi_W&=-4\pi G \frac{a^2}{k^2}\bar{\rho}_\m\delta \label{eq:WeylCQ}\, . 
\end{align}
The evolution equation for $\delta$ can then be obtained by combining the two Euler equations, the two continuity equations and the Poisson equation~\eqref{eq:WeylCQ}:
\begin{align}
\delta''+\left[\HH+\beta x\bar{\varphi}'\frac{\delta'_\dm}{\delta'}\right]\delta'-4\pi G a^2 \bar{\rho}_\m  
\left[1+\frac{\beta^2 x^2}{4\pi G}\frac{\delta_\dm}{\delta}\right]\delta=0 \, . \label{eq:evolCQ}  
\end{align}
Eq.~\eqref{eq:evolCQ} cannot be solved on its own since it depends on $\delta_\dm$. For a given model, we can however find the solution for $\delta_\dm$ from the system of equations~\eqref{eq:00CQ} to~\eqref{eq:EulerCQb}, and then insert this solution into~\eqref{eq:evolCQ} to find $\delta$.\footnote{Equivalently, we could find the solution for $\delta_\dm$ and $\delta_\ba$ from the system of equations, and then compute $\delta=x\delta_\dm+(1-x)\delta_\ba$. However, passing through Eq.~\eqref{eq:evolCQ} makes the degeneracies between the different models clearer.} The solution can then be compared with the measured $V_\g$ and $\Psi_W$ using Eqs.~\eqref{eq:contgCQ} and~\eqref{eq:WeylCQ}.

Here we have assumed that $x$ is constant over time. In reality $x$ may vary, since the interaction of dark matter with dark energy can modify the evolution of $\bar{\rho}_\dm$, which may not decay as $a^{-3}$ anymore. Such deviations in the background evolution are however tightly constrained, since they directly impact the expansion rate of the Universe. In any case, accounting for the variation of $x$ has no impact on the degeneracy argument and on the outcome of the three tests. It would induce a modification in the continuity equation for the galaxy density~\eqref{eq:contgCQ}, and generate additional modifications in the evolution equation~\eqref{eq:evolCQ} and the Euler equation~\eqref{eq:EulerCQ}. The outcome of the three tests would, however, remain unchanged: $\testij$ would still vanish, while $\testzi$ and $\testeuler$ would remain non-zero, yielding exactly the same outcome as for constant $x$.

\subsection{Viscous dark matter}
\label{app:vDM}

The energy-momentum tensor of dark matter is given by Eq.~\eqref{eq:TvDM}. The linearized system of equation in the QSA takes the form 
\begin{align}
&{\rm Einstein\ 00\ (Poisson)}: k^2\Phi=-4\pi G a^2 \delta\rho\, , \label{eq:00vdm}\\   
&{\rm Einstein\ } i\neq j: \Phi-\Psi=-4\pi G a^2\frac{\ep \bar{\rho}_\dm}{\HH}\frac{V_\dm}{k}\, ,  \label{eq:ijvdm}\\
&{\rm Einstein\ } 0i: \Phi'+\HH\Psi=-\frac{4\pi G a^2}{k}\left[\bar{\rho}_\ba V_\ba+(1-\Z)\bar{\rho}_\dm V_\dm \right]\, ,\label{eq:0ivdm}\\
&{\rm Continuity\ dm}: \delta'_{\rm dm}=(1-2\Z)k V_{\rm dm}-3\HH\Z\delta_\dm\, ,  \label{eq:contvdmdm}\\
&{\rm Continuity\ b}: \delta'_{\rm b}=k V_{\rm b}\, ,  \label{eq:contvdmb}\\
&{\rm Euler\ dm}: V'_{\rm dm}+\HH\left[1+3\Z-\frac{\Z'}{\HH(1-\Z)}+\frac{\ep+\Z}{3(1-\Z)}\left(\frac{k}{\HH}\right)^2\right] V_{\rm dm}+\frac{k\Psi}{1-\Z}+\frac{\Z}{1-\Z}\frac{k\Phi'}{\HH}=0\, , \label{eq:Eulervdmdm}\\
&{\rm Euler\ b}: V'_{\rm b}+\HH V_{\rm b}+k\Psi=0\, .\label{eq:Eulervdmb}
\end{align}
The parameters $\ep$ and $\Z$ are related to the bulk and shear viscosity through Eq.~\eqref{eq:vDM_epzeta}. As already pointed out in~\cite{Zheng:2025oiq}, the correction to the Euler equation is strongly enhanced relative to the corrections to the Einstein and continuity equations. The Euler equation indeed acquires modifications of the form
  \begin{align}
1 \rightarrow 1+ \frac{\ep+\Z}{1-\Z} \left(\frac{k}{\HH}\right)^2\, ,
\end{align}
whereas all other equations are modified as~\footnote{For the gravitational slip equation, we can see this by noting that $4\pi G a^2\bar{\rho}_\dm \sim \HH^2$ and that $V_\dm\sim k/\HH\Phi$ (from Eq.~\eqref{eq:0ivdm}), leading to $\Phi-\Psi\sim \ep \Phi$.}
\begin{align}
1 \rightarrow 1+ \ep \quad \mbox{or}\quad 1 \rightarrow 1+ \Z\, .
\end{align}
For modes well inside the horizon, $k/\HH\gg 1$, and therefore the only relevant modification is the one appearing in the Euler equation. In other words, a detectable modification of the Einstein and continuity equations would imply a much larger correction to the Euler equation, dramatically altering the growth of structure. Such an effect would already have been detected through measurements of $V_\g$ and $\Psi_W$. In the following, we can therefore safely neglect the subdominant contributions and retain only the terms proportional to $(k/\HH)^2$ in Eq.~\eqref{eq:Eulervdmdm}.

Assuming that $x=\bar{\rho}_\dm/\bar{\rho}$ is constant (which is a good approximation since the background continuity equation for dark matter receives only subdominant corrections) the Euler equation for the galaxy velocity becomes
\begin{align}
V_\g'+\HH V_\g+k\Psi=-x\HH\frac{\ep+\Z}{3(1-\Z)}\left(\frac{k}{\HH}\right)^2V_\dm\, .   
\end{align}
The galaxy velocity and the Weyl potential can be related to the matter density through
\begin{align}
kV_{\rm g}&=\delta'\, ,\label{eq:contgvdm}\\
\Psi_W&=-4\pi G \frac{a^2}{k^2}\bar{\rho}_\m\delta\label{eq:Weylvdm}\, .
\end{align}
And the matter density obeys the evolution equation 
\begin{align}
\delta''+\HH\left[1+x\frac{\ep+\Z}{3(1-\Z)}\left(\frac{k}{\HH} \right)^2\frac{\delta'_\dm}{\delta'}\right]\delta'-4\pi G a^2 \bar{\rho}_\m  \delta=0 \, . \label{eq:evolvdm}  
\end{align}

Interestingly, the fact that the Euler equation acquires a correction enhanced by a factor $(k/\HH)^2$ relative to those in the gravitational slip equation and in the $0i$ Einstein equation remains true for dark matter models with a generic anisotropic stress $\Pi^i_j=(\partial^i\partial_j-\frac{1}{3}\delta^i_j\nabla^2)\Pi$. In this case the gravitational slip equation is indeed modified as 
\begin{align}
\Phi-\Psi=8\pi G a^2 \Pi\, .  \label{eq:ijanis} 
\end{align} 
The $0i$ Einstein equation and the continuity equations are not modified, while the Euler equation becomes
\begin{align}
V'_{\rm dm}+\HH V_{\rm dm}+k\Psi=-\frac{2}{3}k^3\frac{\Pi}{\bar{\rho}_{\rm dm}}\, . \label{eq:Euleranis}
\end{align}
From Eq.~\eqref{eq:ijanis} we have that $\Pi/\bar{\rho}_{\rm dm}\sim (\Phi-\Psi)/\HH^2$. Inserting this in Eq.~\eqref{eq:Euleranis}, we find that the right-hand side of the Euler equation is of the order of $(k/\HH)^2k(\Phi-\Psi)$. This means that a correction of order unity in the Euler equation~\eqref{eq:Euleranis} leads to a correction of order $(\HH/k)^2\ll 1$ in the gravitational slip equation~\eqref{eq:ijanis}. Consequently, dark matter models with non-vanishing anisotropic stress lead to $\testzi=\testij=0$ and $\testeuler\neq 0$. They would therefore be correctly identified as dark matter models exhibiting deviations from perfect-fluid behaviour.

\subsection{Evolving dark energy}
\label{app:eDE}

The action for evolving dark energy is given in Eq.~\eqref{eq:SeDE}. Since dark energy does not cluster, it does not contribute to the Einstein equations. The continuity and Euler equations are also unaffected by the scalar field, since it is minimally coupled to gravity. The linearized equations keep therefore the same form as in $\Lambda$CDM
\begin{align}
&{\rm Einstein\ 00\ (Poisson)}: k^2\Phi=-4\pi G a^2 \delta\rho\, , \label{eq:00ede}\\   
&{\rm Einstein\ } i\neq j: \Phi-\Psi=0\, , \label{eq:ijede}\\
&{\rm Einstein\ } 0i: \Phi'+\HH\Psi=-4\pi G a^2\bar{\rho}_\m\frac{V}{k}\, ,\label{eq:0iede}\\
&{\rm Continuity}: \delta'=k V\, , \label{eq:contede}\\
&{\rm Euler}: V'+\HH V+k\Psi=0\, . \label{eq:Eulerede}\\
\end{align}
The Euler and continuity equations involving $V_\g$ are standard, as well as the Poisson equation for $\Psi_W$ and the evolution equation for $\delta$ 
\begin{align}
&kV_{\rm g}=\delta'\, ,\label{eq:contgede}\\
&V_\g'+\HH V_\g+k\Psi=0\, ,\\
&\Psi_W=-4\pi G \frac{a^2}{k^2}\bar{\rho}_\m\delta\label{eq:Weylede}\, .\\
&\delta''+\HH\delta'-4\pi G a^2 \bar{\rho}_\m\delta=0\, .  \label{eq:evolede}
\end{align}
The difference in the growth of $\delta$ with respect to $\Lambda$CDM is only due to $\bar{\rho}_\de$, which modifies the evolution of $\HH$. This impacts $\delta$ through the second term in Eq.~\eqref{eq:evolede} and consequently it impacts also $V_\g$ and $\Psi_W$.

\subsection{Clustering dark energy}
\label{app:cDE}

The action for clustering dark energy is given in Eq.~\eqref{eq:ScDE}. Contrary to the smooth case of Section~\ref{app:eDE}, here dark energy acquires perturbations, such that $\delta\rho=\delta\rho_\m+\delta\rho_\de$ and $V_\de\neq 0$. The system of linearized equations for models with negligible sound speed takes the form~\cite{Sefusatti:2011cm}:
\begin{align}
&{\rm Einstein\ 00\ (Poisson)}: k^2\Phi=-4\pi G a^2 (\delta\rho_{\rm m} +\delta\rho_{\rm de})\, ,\label{eq:00cde}\\   
&{\rm Einstein\ } i\neq j: \Phi-\Psi=0 \, ,\label{eq:ijcde}\\
&{\rm Einstein\ } 0i: \Phi'+\HH\Psi=4\pi G a^2\left(\bar{\rho}_\de(1+w_\de)\frac{\delta\varphi}{\bar{\varphi}'}-\bar{\rho}_\m\frac{V_\m}{k} \right)\, ,\label{eq:0icde}\\
&{\rm Continuity\ m}: \delta'_\m=k V_\m \, ,\label{eq:contcdem}\\
&{\rm Continuity\ de}: \delta'_\de-3\HH w_\de\delta_\de=k(1+w_\de) V_\de\, , \label{eq:contcdede}\\
&{\rm Euler\ m}: V'_\m+\HH V_\m+k\Psi=0 \, ,\label{eq:Eulercdem}\\
&{\rm Euler\ de}:  V'_\de+\HH V_\de+k\Psi=0 \, .\label{eq:Eulercdede}
\end{align}
The Euler equations for matter and for dark energy are the same and the two fluids are therefore comoving: $V_\m=V_\de=V_\g$. Hence the Euler equation for galaxies is standard
\begin{align}
&V_\g'+\HH V_\g+k\Psi=0\, . \label{eq:eulergcde}    
\end{align}
The continuity equation for the total density field~\footnote{Note that here we use the same convention as in~\cite{Sefusatti:2011cm} and we define the density contrast with respect to $\bar{\rho}_\m$ in the denominator, since it is convenient. We could alternatively define $\delta$ with $\bar{\rho}=\bar{\rho}_\m+\bar{\rho}_\de$ in the denominator. This is a choice of definition, which changes the form of the equations but does not impact the physics. In particular it does not change the form of the three tests, which do not rely on $\delta$.}
\begin{align}
\delta=\frac{\delta\rho_\m+\delta\rho_\de}{\bar{\rho}_\m}    
\end{align}
is however modified. Combining Eqs.~\eqref{eq:contcdem} and~\eqref{eq:contcdede} we find
\begin{align}
kV_{\rm g}&=\left(\frac{\delta'_\m}{\delta'}\right)\delta'\, .\label{eq:contgcde}
\end{align}
The Poisson equation for the Weyl potential takes the form
\begin{align}
\Psi_W&=-4\pi G \frac{a^2}{k^2}\bar{\rho}_\m\delta \label{eq:Weylcde}\, .
\end{align}
Finally, combining the continuity equations, Euler equations and the Poisson equation, we obtain an evolution equation for $\delta$
\begin{align}
\delta''&+\left\{\big[\HH(1+x_\de(1+ w_\de) )-x'_\de w_\de- x_\de w'_\de  \big]\frac{\delta'_\m}{\delta'} -\big[3\HH x_\de w_\de +x'_\de \big]\frac{\delta'_\de}{\delta'} \right\}\delta'\label{eq:evolcde}\\
&-\left\{
\big[3\HH x_\de w_\de +x_\de' \big]'\frac{\delta_\de}{\delta} +4\pi G a^2\bar{\rho}_\m\big[1+x_\de(1+ w_\de) \big] \right\}\delta=0\, ,  \nonumber
\end{align}
where $x_\de\equiv \bar{\rho}_\de/\bar{\rho}_\m$. Note that here we cannot assume $x_\de$ to be constant over time, since dark energy evolves in a drastically different way than matter. As for CQ, to solve for $\delta$, we first need to solve the coupled system of equations~\eqref{eq:00cde} to~\eqref{eq:Eulercdede} to find $\delta_\de$ and $\delta_\m$. Eq.~\eqref{eq:evolcde} is however useful to see the similarities with the other models.

\section{Recovering the $0i$ Einstein equation from the other equations in GBD}

To illustrate the redundancy of the $0i$ Einstein equation when we restrict the space of theories to modified gravity only, we show how this equation can be recovered in GBD by combining the Poisson and gravitational slip equations with the Klein-Gordon equation and the continuity equation. 

We first take the derivative of the Poisson equation~\eqref{eq:00GBD}
\begin{align}
\Phi'=-\frac{4\pi G A^2 a^2}{k^2}\left[2\HH\bar{\rho}_\m+2\beta\bar{\varphi}'\bar{\rho}_\m +\bar{\rho}'_m \right]\delta 
-\frac{4\pi G A^2 a^2}{k^2}\bar{\rho}_\m \delta'-\beta\delta\varphi'-\beta_{,\varphi}\bar{\varphi}'\delta\varphi\, .
\end{align}
Using Eq.~\eqref{eq:KGGBD} to rewrite $\delta$ in terms of $\delta\varphi$ and Eq.~\eqref{eq:EulerGBD} to rewrite $\delta'$ in terms of $V$, and using that $\bar{\rho}'_\m=-3\HH\rho_\m$, we obtain
\begin{align}
\Phi'=\left[4\pi G A^2\left(2\bar{\varphi}'-\frac{\HH}{\beta} \right) -3\beta\HH+6\beta^2\bar{\varphi}'-\beta_{,\varphi}\bar{\varphi}'\right]\delta\varphi-\beta\delta\varphi'-4\pi G A^2 a^2\bar{\rho}_\m\frac{V}{k}\, .    
\end{align}
Similarly, combining Eq.~\eqref{eq:PoissonGBD} with Klein-Gordon equation, we can rewrite $\Psi$ in terms of $\varphi$
\begin{align}
\Psi=\left[4\beta+\frac{4\pi G A^2}{\beta}\right]\delta\varphi\, . \label{eq:PsivarphiGBD}  
\end{align}
From this we obtain
\begin{align}
\Phi'+\HH\Psi=-4\pi Ga^2A^2\bar{\rho}_\m\frac{V}{k}-\beta\delta\varphi'+\left[8\pi G A^2\bar{\varphi}'+\beta\HH+6\beta^2\bar{\varphi}'-\beta_{,\varphi}\bar{\varphi}'\right]\delta\varphi\, .  
\end{align}
Subtracting $-\beta\bar{\varphi}'\Psi$ on both side of the equation and using~\eqref{eq:PsivarphiGBD} on the right-hand side we obtain the $0i$ Einstein equation~\eqref{eq:0iGBD}.  

\bibliographystyle{apsrev4-2}
\bibliography{Einstein0i}

\end{document}